\documentclass[preprint2]{aastex6}
\begin{document}
\title{Failed Growth at the Bouncing Barrier in Planetesimal Formation}
\author{Maximilian Kruss, Tunahan Demirci, Marc Koester, Thorben Kelling, Gerhard Wurm}
\affil{Faculty of Physics, University Duisburg-Essen, Lotharstr. 1, 47057 Duisburg, Germany; marc.koester@uni-due.de}

\begin{abstract}
In laboratory experiments, we studied collisions of ensembles of compact (filling factor 0.33) millimeter dust aggregates composed of micrometer quartz grains. We used cylindrical aggregates, triangular aggregates, square aggregates, and rectangular aggregates. Ensembles of equal size aggregates as well as ensembles with embedded larger aggregates were studied. The typical collision velocities are 10--20~mm~s$^{-1}$. High spatial and temporal resolution imaging unambiguously shows that individual collisions lead to sticking with a high probability of 20\%. This leads to connected clusters of aggregates. The contact areas between two aggregates increase with collision velocity. However, this cluster growth is only temporary, as subsequent collisions of aggregates and clusters eventually lead to the detachment of all aggregates from a cluster. The contacts are very fragile as aggregates cannot be compressed further or fragment under our experimental conditions to enhance the contact stability. Therefore, the evolution of the ensemble always leads back to a distribution of individual aggregates of initial size. This supports and extends earlier experiments showing that a bouncing barrier in planetesimal formation would be robust against shape and size variations.
\end{abstract}

\maketitle

\section{Introduction}
\label{sec:intro}

This work is focused on a process, discussed in the context of the early phases of planetesimal formation, known as the bouncing barrier \citep{Zsom2010}. Initially, dust grains in protoplanetary disks collide, stick together and grow \citep{Dominik1997, Wurm1998, Blum2008, Wada2009}. The aggregates are initially fractal or very porous and if they consist of sub-micron icy grains they might grow large in the above fashion \citep{Okuzumi2012, Kataoka2013}. However, if they consist of micrometer sized silicates they are supposed to become compact at millimeter to centimeter sizes. 
\citet{Weidling2009} showed that an initially highly porous aggregate with a volume filling factor of 0.15 has a volume filling factor of 0.36 after many collisions with a wall at velocities up to 0.35~m~s$^{-1}$. Experiments by \citet{Teiser2011} and \citet{Meisner2012} showed that a filling factor of 0.3--0.4 seems to be a common value in collisional evolution up to high speeds of several tens of meters per second for (sub-) millimeter aggregates. 
In any case, aggregates are eventually compacted to such a level that further collisions no longer include large dissipations of energy through restructuring the whole aggregate. The aggregates then tend to bounce off each other in mutual collisions, as observed in many individual experiments. This has also triggered theoretical work on understanding bouncing \citep{Wada2011, Seizinger2013}, which supports that growth no longer proceeds under the premise of continuous bouncing. This barrier was introduced by \citet{Zsom2010} as the bouncing barrier. 

The existence of a bouncing barrier has severe consequences for planetesimal formation. If the aggregate size where bouncing dominates and growth becomes stalled is too small, pure growth of planetesimals cannot proceed. If larger seeds are provided by some mechanism, the growth of larger bodies is possible by mass transfer \citep{Teiser2009, Windmark2012a, Deckers2014}. In that case the bouncing barrier is required to keep high the number of small aggregates on which the larger seeds can feed. Alternatively, large objects can form by gravitational collapse of clumps of particles concentrated previously e.g. in zonal flows or by streaming instabilities \citep{Dittrich2013, Johansen2014}. Early work mostly considered particles to be concentrated by streaming instability if their Stokes numbers (the ratio between gas--grain coupling time and orbital period) are on the order of 1. However, recent work showed that much lower Stokes numbers might suffice. \citet{Bai2010} find the onset of a streaming instability for Stokes numbers larger than $10^{-2}$. This is consistent with work by \citet{Carrera2015} who find the onset of streaming instability at Stokes numbers between $10^{-3}$ and $10^{-2}$. In this work this is equivalent to millimeter-sized particles and we also consider the possibility that collisions of millimeter-sized particles (chondrules or chondrule aggregates) at low speeds might lead to further growth, enhancing the effect of the streaming instability. \citet{Drazkowska2014} combine a model for streaming instability and particle coagulation, and show that the bouncing barrier for silicate dust might be below the limit required for streaming to set in (Stokes number $10^{-2}$). In general, a particle reservoir at the bouncing barrier (often referred to as pebbles) might benefit accretion by larger bodies \citep{Johansen2015}. 

Due to the importance of the bouncing barrier, we extend earlier laboratory experiments on this topic.
\citet{Kelling2014} supported the existence of a bouncing barrier. They studied ensembles of compact aggregates interacting with each other in thousands of collisions per aggregate. 
They found no long-term growth of a larger cluster of aggregates.
In those earlier experiments the sticking probability in individual collisions could not be determined unambiguously as the spatial and time resolutions were not sufficient. Also, only one size and shape of
initial aggregates was used. To elaborate more on the robustness of the bouncing barrier we improve the
set-up and extend the parameter set studied.

A high resolution camera system was installed to allow the study of individual collisions. We now can clearly classify individual collisions in terms of bouncing or sticking and quantify contact areas. Also, various aggregate shapes were studied. The reason behind the choice of triangular- and square-shaped aggregates is that sticking at long sides might form stronger bonds. This might enhance the stability of clusters in comparison to 
cylindrical aggregate clusters where curvature would, e.g., prevent sticking at several distinct sides of two aggregates. We do not expect aggregates in protoplanetary disks to be such shapes. In fact, if aggregates grow and are compacted by colliding with other grains from all sides they will tend to be roundish. However, this depends on the overall distribution of solids within the disk, e.g., a fraction of large bodies might exist at the same time (as needed in pebble accretion scenarios, \citet{Johansen2015}). Fragments of their collisions might join the sub-bouncing barrier particle fraction. Then aggregates might not be spherical but have angular shapes. So our choice is meant to evaluate the possible extremes that might break the bouncing barrier. In addition, the ensemble evolution was studied with individual larger aggregates embedded. The idea here is that larger aggregates might act as seeds for more stable clusters.
In summary, we consider the influence of (1) aggregate shape and (2) aggregate size differences, and (3) quantify the underlying fraction of individual sticking collisions as necessary for growth.\\

\section{Experimental set-up}
\label{sec:setup}

\begin{figure}
\includegraphics[width=\columnwidth]{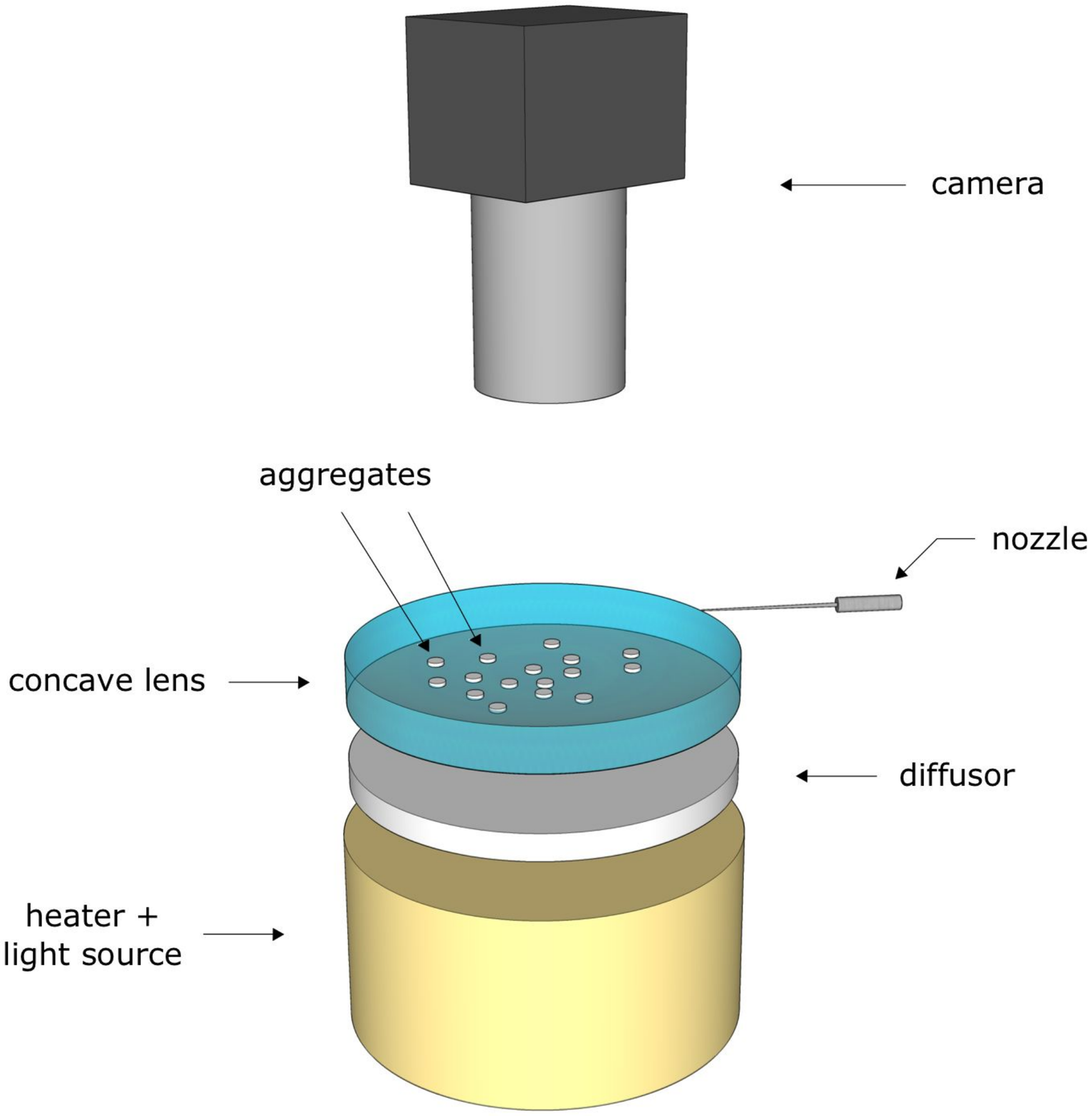}
    \caption{\label{fig.setup1}Principles of the basic set-up of the experiment; A concave lens is heated and illuminated diffusely from below. Dust aggregates on top levitate under low
    pressure conditions (vacuum chamber not shown). The motion is observed by
    a camera outside the vacuum chamber. Aggregates always move in random motion due to asymmetries in the levitation. A nozzle can be used to excite the velocities.}
\end{figure}

The principle experimental set-up is shown in Figure \ref{fig.setup1}. 
Dust aggregates are placed on a back-lit surface heated to 900 K. At an ambient pressure of 15 mbar the dust aggregates lift off and hover over the surface due to thermal creep \citep{Kelling2009}. 
The free aggregates have random speeds which lead to
collisions. The speed can be enhanced by exciting the aggregates with a short air pulse.
Compared to the forces during collisions, the remaining gravitational force of the slightly curved heater
and the mostly repulsive gas drag associated with the lifting principle are negligible. 
For further details we refer the reader to \citet{Kelling2014}. It should be noted here, however, that in this earlier work the low spatial and temporal resolutions did not allow the unambiguous qualification of individual collisions as being bouncing or being repelled by the gas flow without actual collisions.

\begin{figure}
\includegraphics[width=\columnwidth]{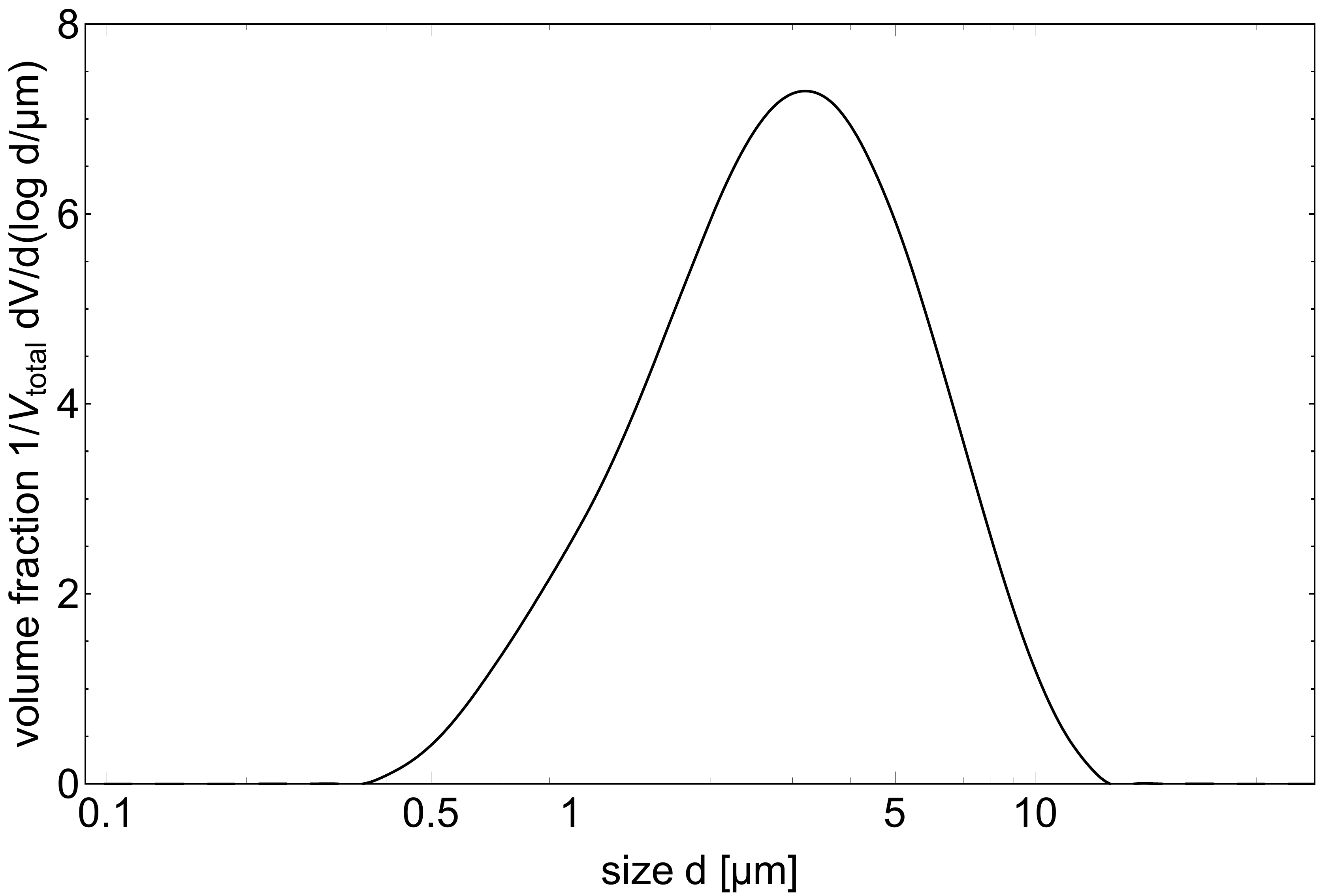}
    \caption{\label{fig.sizedistribution}Size distribution of the quartz grains as basic entities of the dust aggregates; The distribution was measured by a commercial
    instrument based on light scattering (Malvern Mastersizer 3000).}
\end{figure}

\begin{figure}
\includegraphics[width=\columnwidth]{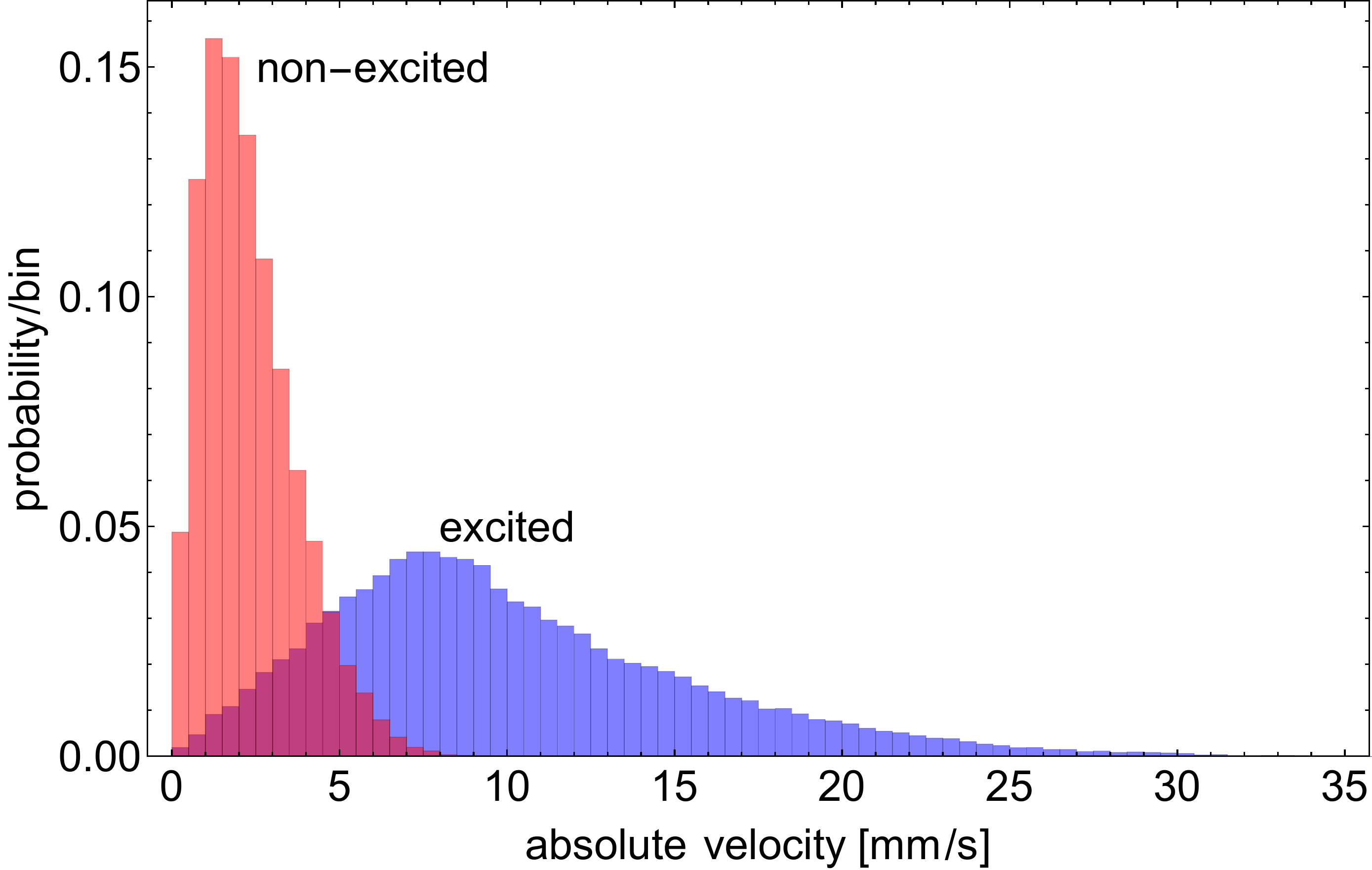}
    \caption{\label{fig.velocities}Example of the absolute velocity distribution of eight aggregates within an ensemble tracked for 4.3 s; red: without excitation; blue: with excitation}
\end{figure}

As in the earlier work, dust aggregates are manually pressed and removed from molds, but these are now of different size and shape. We have three different scales in the experiment, the dust grains, the aggregates built from the dust grains, and (small) clusters built from the aggregates. The basic dust grains are irregular $\rm SiO_2$ quartz grains with a size range between 0.1 and 10 $\rm \mu m$. A detailed size distribution is shown in Figure \ref{fig.sizedistribution}.
The height of the dust aggregates is determined by the thickness of the molds and is always 0.2 mm.
Sample dust aggregates have been weighed and were found to have volume filling factors of 0.33. This agrees with the natural value to be expected in collisions, as quoted above \citep{Teiser2011, Meisner2012}.
At low spatial resolution the absolute velocity of the aggregates was tracked and gives the velocity distributions shown in Figure \ref{fig.velocities}. 
This indicates the order of magnitude and distributions of the collision velocities (quantified below) to be expected for collisions.\\

\section{Results}
\label{sec:results}

\subsection{Individual collisions}

\begin{figure}
\includegraphics[width=\columnwidth]{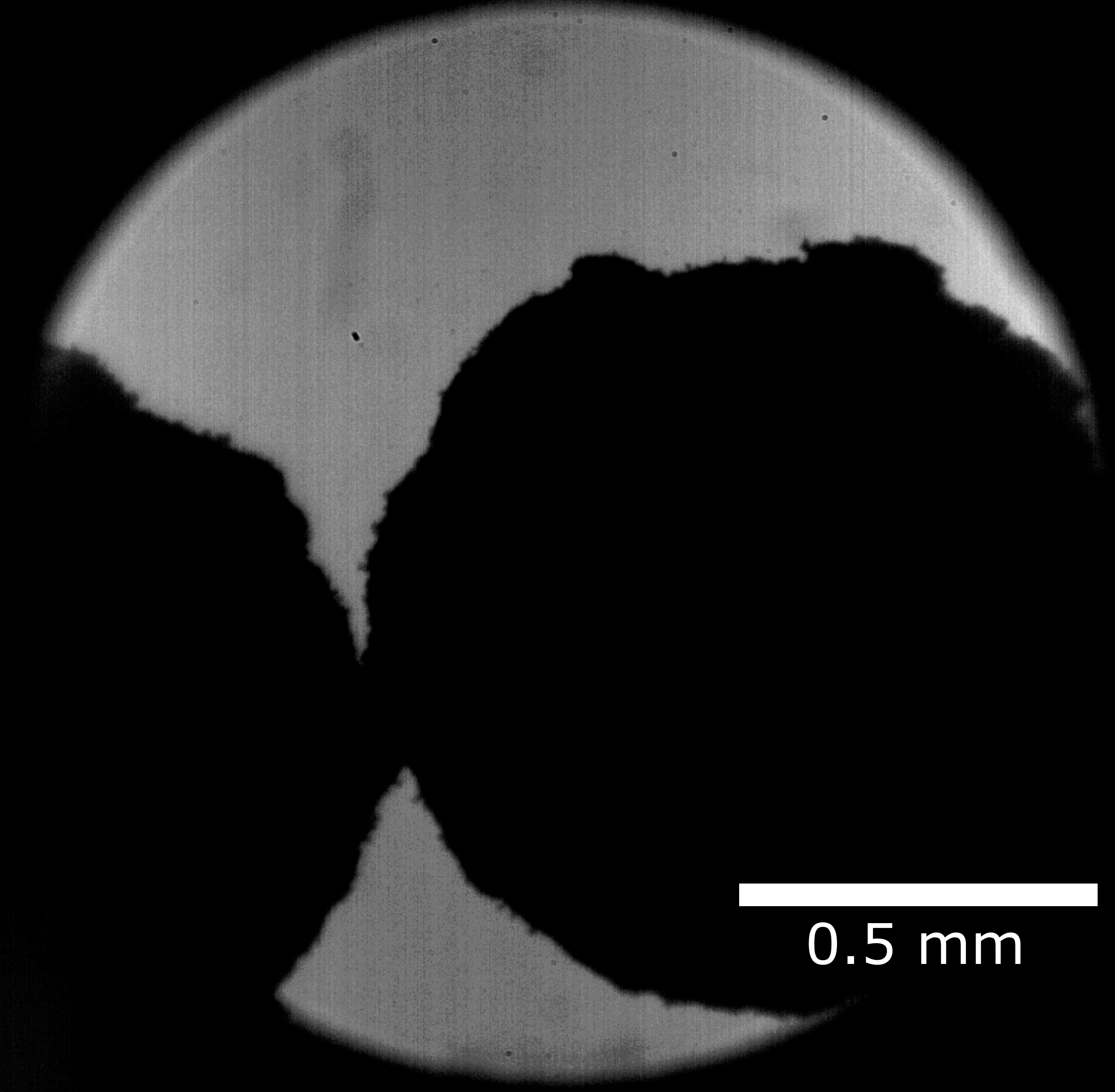}
    \caption{\label{fig.collidesample}Example image of two aggregates sticking after a collision. The bright circle is the field of view.}
\end{figure}

\begin{figure}
\includegraphics[width=0.982\columnwidth]{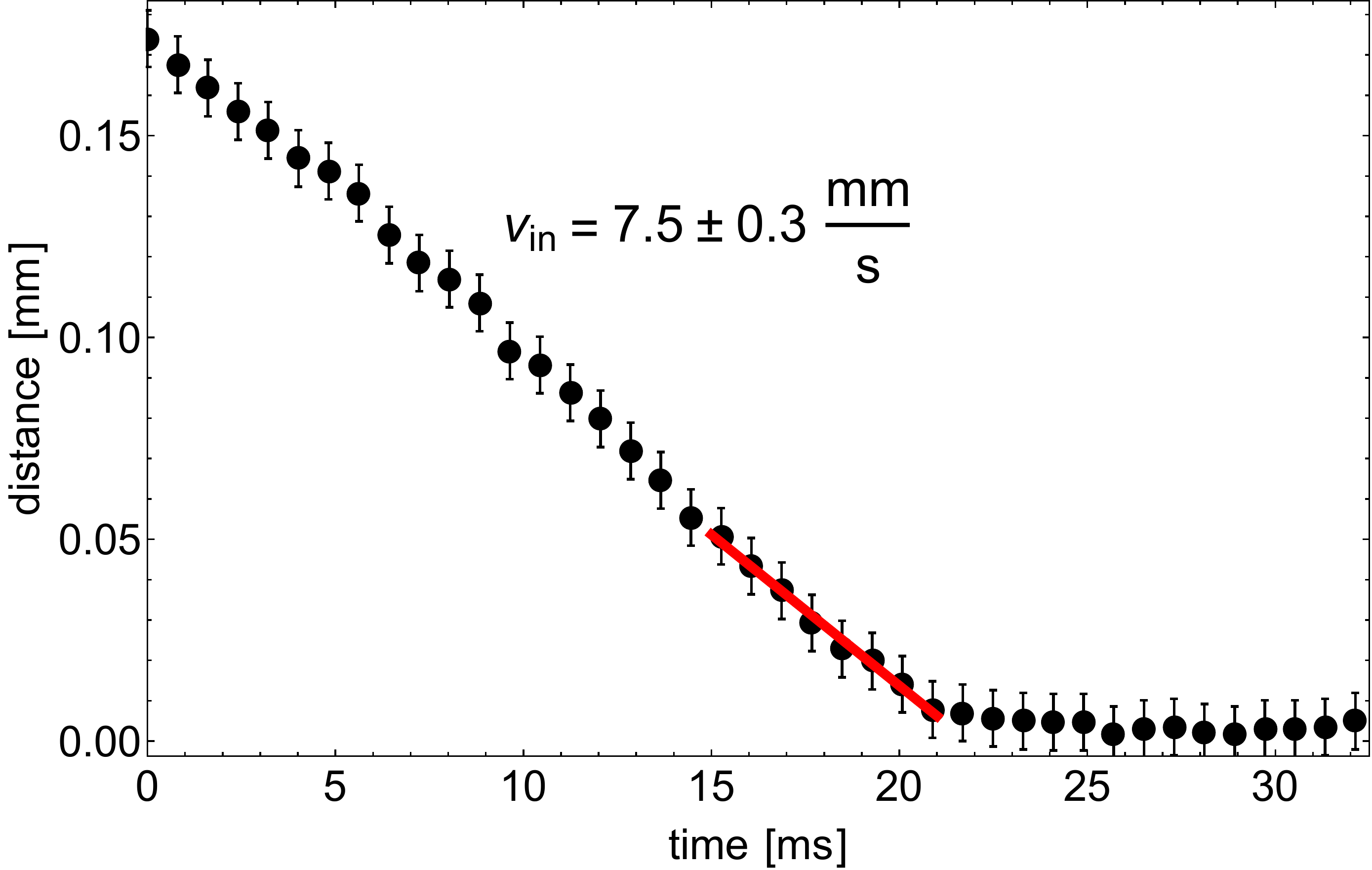}
\includegraphics[width=\columnwidth]{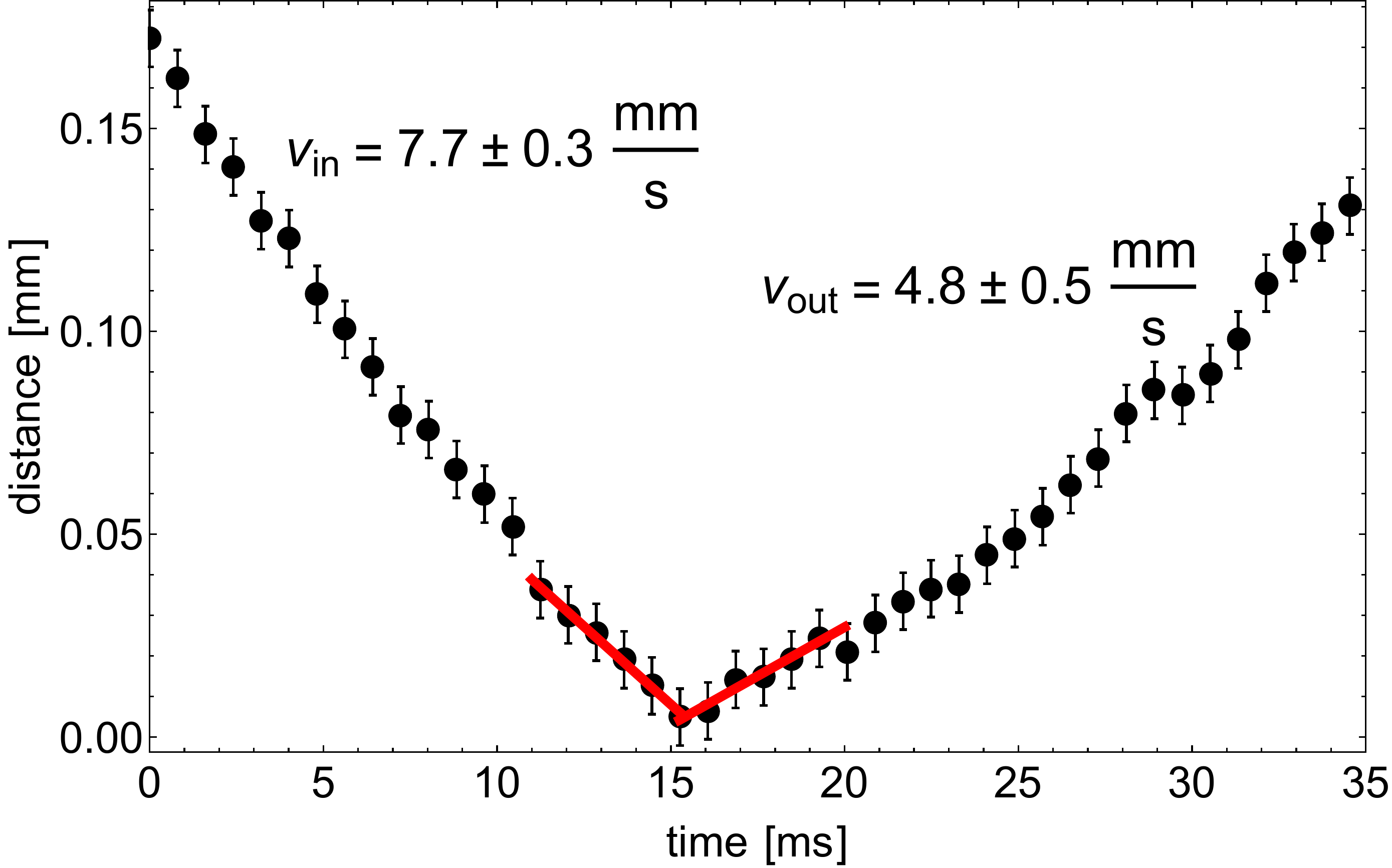}
\caption{\label{fig.distance}Distance over time for a sticking collision (top) and bouncing collision (bottom); the red lines are fits to determine the velocity during approach $v_{\text{in}}$ (collision velocity) and rebound $v_{\text{out}}$.}
\end{figure}

\begin{figure}
\includegraphics[width=\columnwidth]{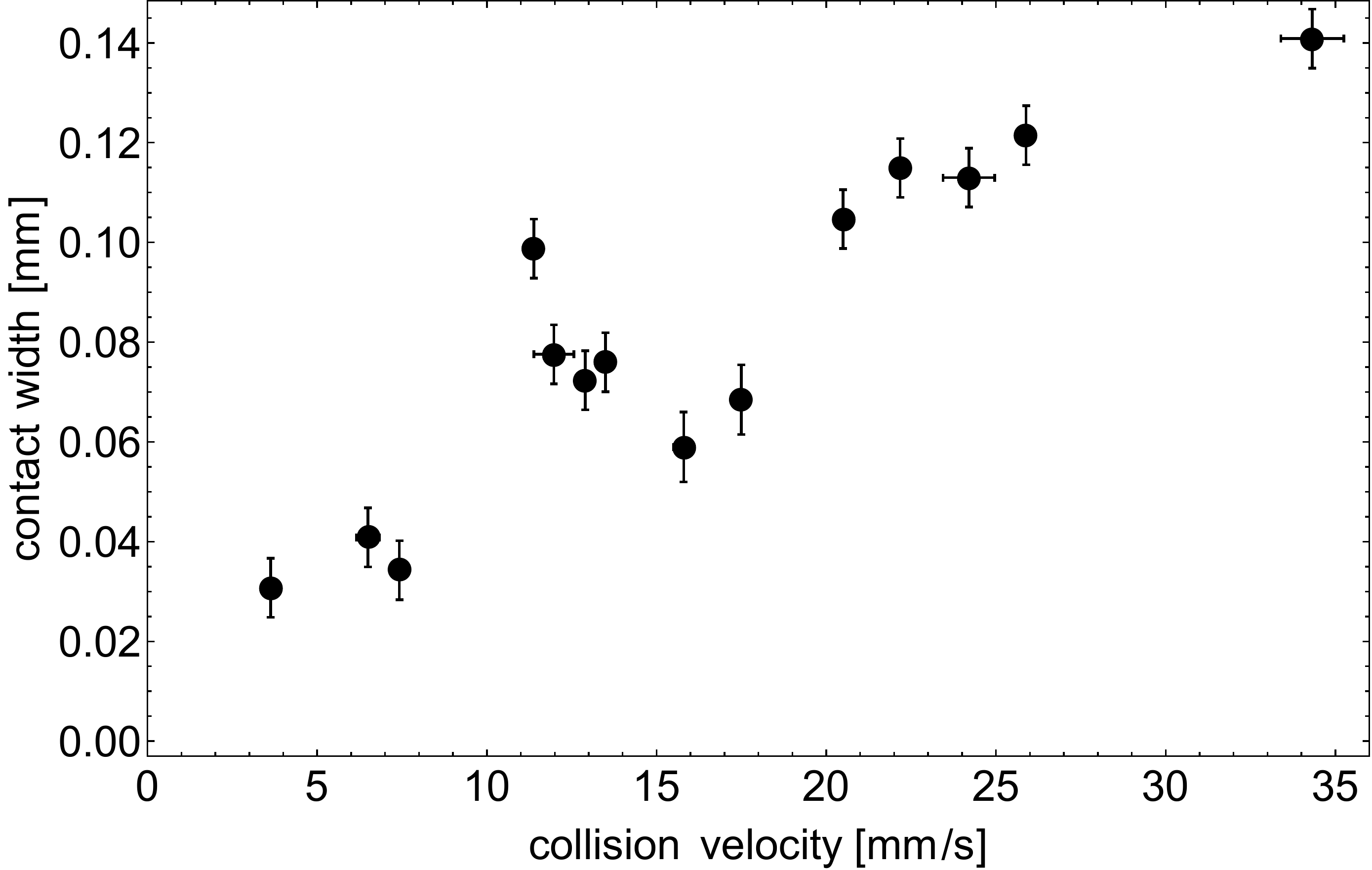}
    \caption{\label{fig.connectingline}Dependence of the contact line width with collision velocity (error bars are partly smaller than the symbol)}
\end{figure}

\begin{figure}
\begin{center}
\includegraphics[width=0.9\columnwidth]{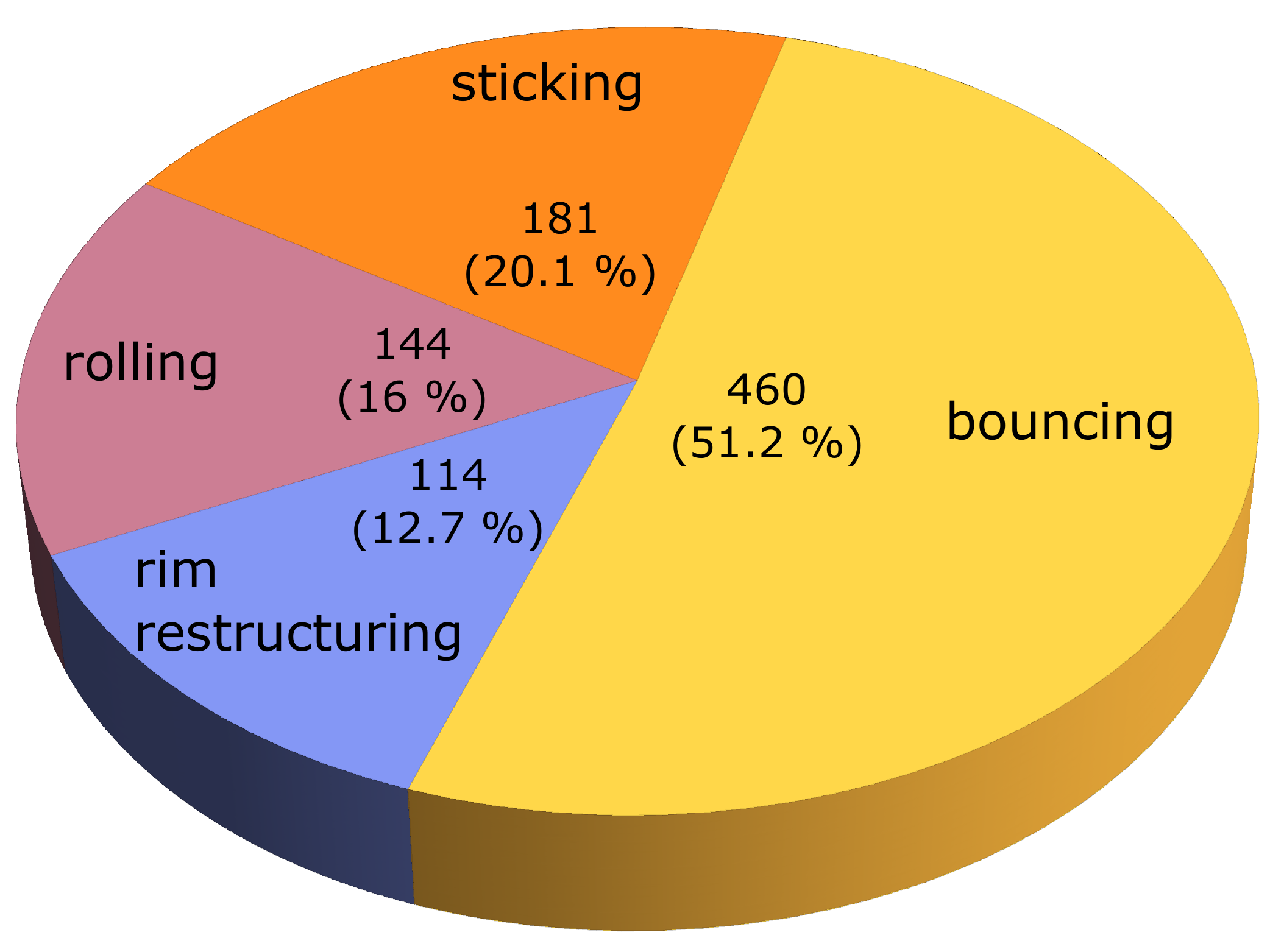}
\end{center}
    \caption{\label{fig.pieceofcake}Fractions of different categories of collisions for a cylindrical aggregate ensemble of 1 mm aggregates. Measured collision velocities are below 35~mm~s$^{-1}$; the absolute numbers observed and percentage of total collisions are given.}
\end{figure}

For cylindrical aggregates of the same size (1 mm in diameter) individual collisions were studied. Collisions were observed at high spatial (1.6 $\rm \mu m$) and temporal resolutions (1000 fps). An example of two aggregates sticking together after a collision is shown in Figure \ref{fig.collidesample}.
From the distance profile with time as seen in Figure \ref{fig.distance}, sticking collisions and bouncing collisions can easily be qualified as such. In addition to the simple hit-and-stick and bouncing collisions we qualitatively distinguish two more sub-types.
In some sticking events the aggregates roll over each other, i.e. the contact point changes while the aggregates stay in contact. Some bouncing collisions show variations of the aggregate rim which we call restructuring. In some events the restructuring looks like mass being transferred from one aggregate to the other.
For the sticking collisions we further determined the extension of the connecting area (a line in the 2D projection). As shown in Figure \ref{fig.connectingline}, there is a clear trend
that higher collision velocities are correlated to larger connecting lines. While the aggregates can be compact with a filling factor of 0.33, the rim obviously still allows some restructuring. As the impact energy has to be dissipated, this small scale restructuring might account for this. It is likely that the outer part of an aggregate continuously changes slightly due to collisions. However, the resulting dimer after a sticking collision of two aggregates is always peanut shaped and the contact between aggregates is a weak point of a cluster of aggregates.
In total, of the 899 collisions studied, 181 lead to sticking and 460 to bouncing.
A detailed distribution of the different kinds of collisions is shown in Figure \ref{fig.pieceofcake}.
Due to the lifting mechanism there is some small
repulsive force, and in earlier work by \citet{Kelling2014} collisions were not resolved enough to rule out that aggregates only approach each other without colliding physically. Our results with the improved imaging now clearly prove the earlier assumption that a large fraction of the collisions leads to sticking. This is a firm result. 
Sticking occurs at velocities of at least up to 35~mm~s$^{-1}$. Due to the bias of the small field of view in high resolution, very small and
very large velocities cannot be quantified. We therefore do not provide a velocity dependent sticking probability.

\subsection{Ensemble evolution}

\begin{figure}
\includegraphics[width=\columnwidth]{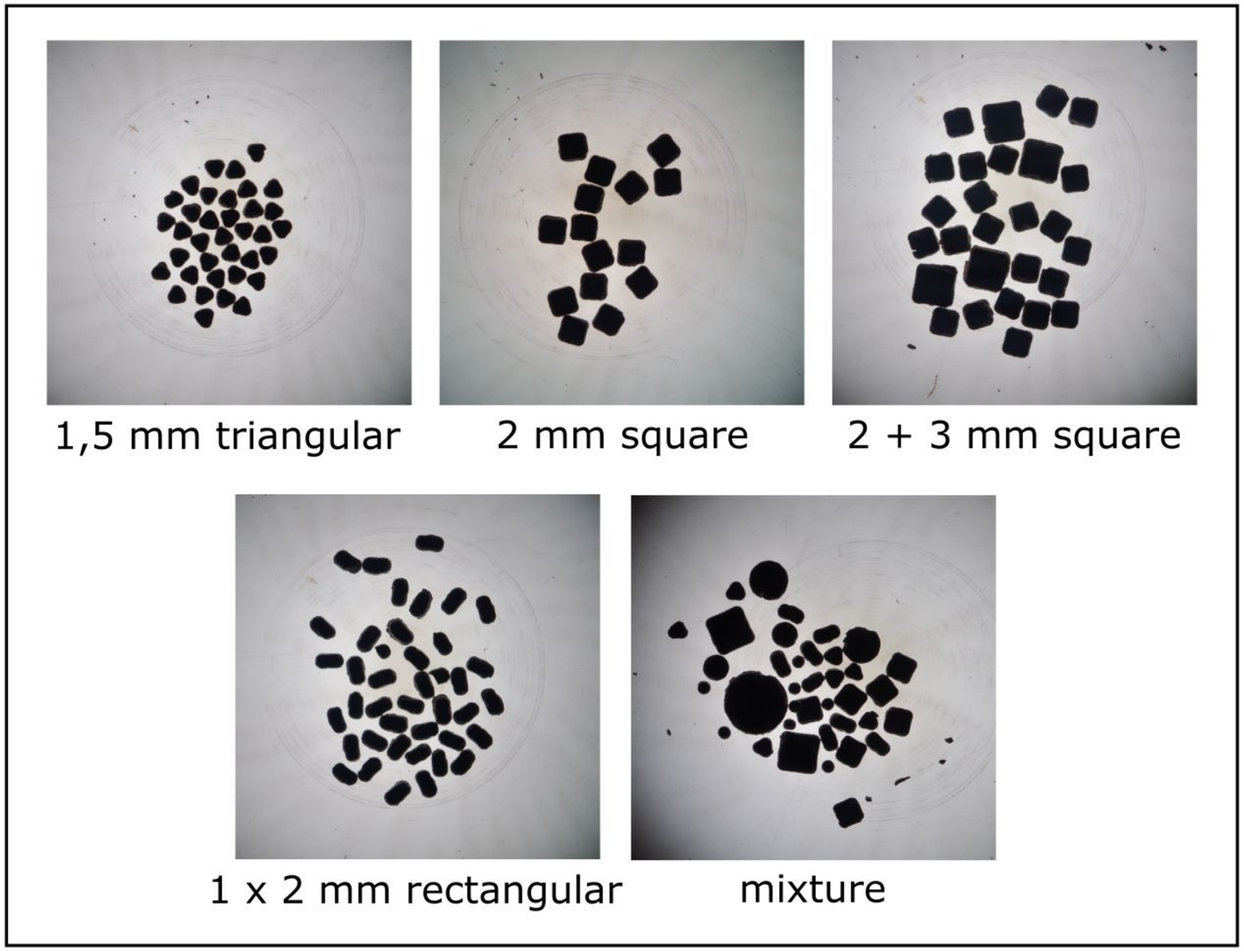}
\includegraphics[width=\columnwidth]{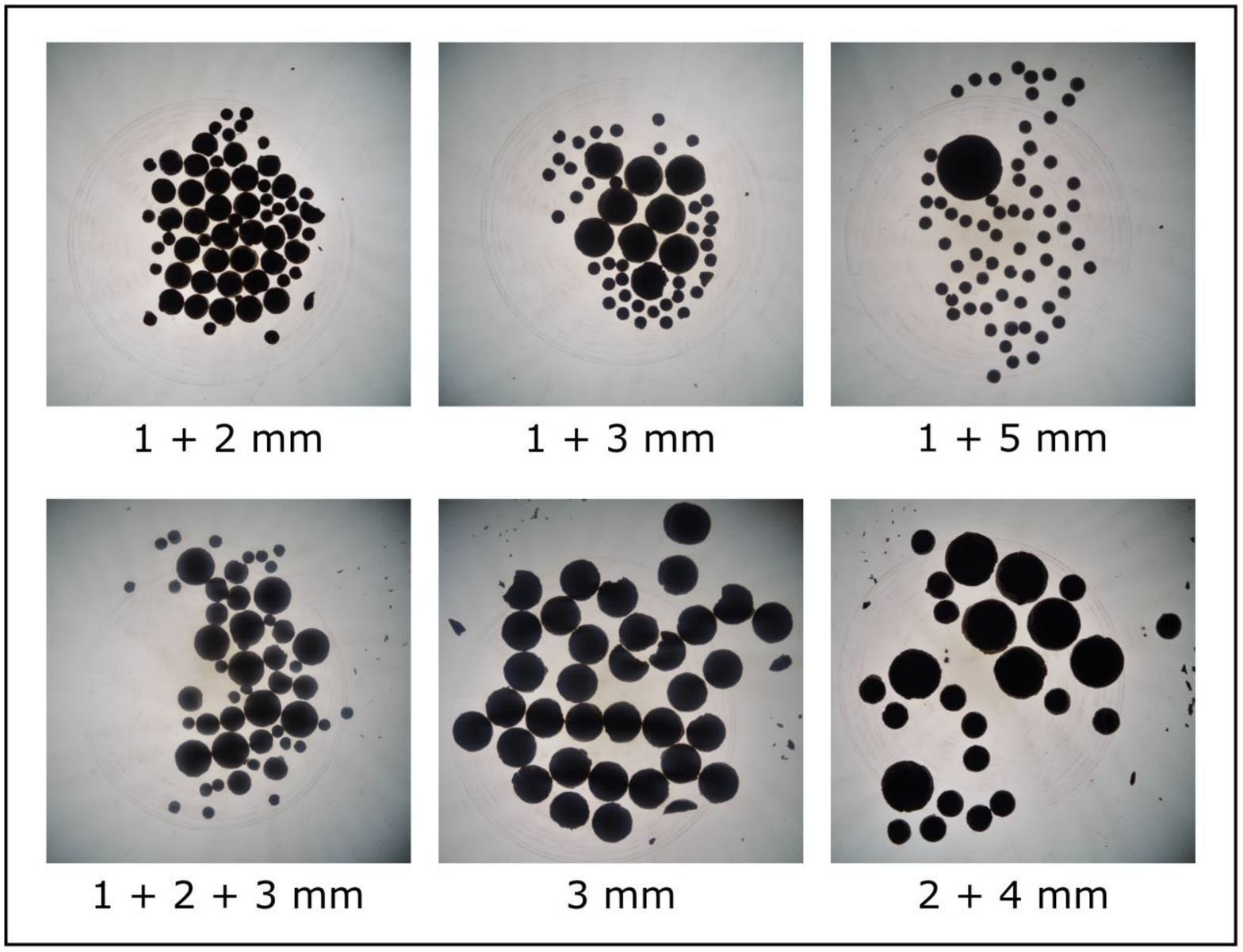}
\caption{\label{fig.variety}The different configurations of aggregate ensembles studied}
\end{figure}

The weak aggregate connections after sticking behave like rated break points and further collisions can separate aggregates within clusters again. To study if specific configurations would allow cluster growth nonetheless, we observed the long-term behavior of collisions
in the aggregate ensembles. Figure \ref{fig.variety} shows different initial settings with different shapes and size ratios.
The average collision velocities for each ensemble are shown in Table \ref{table}.

\begin{table}[b]
\caption{Average Collision Velocities and Standard Deviations for the Different Ensembles Studied}
\vspace{0.3cm}
\begin{tabular}{l|c|c}
Ensemble&$v_{\text{col}}$ (mm s$^{-1}$)&$\sigma$ (mm s$^{-1}$)\\
\hline 
3 mm &24.6&21.1\\
1 + 2 mm& 8.7& 9.2\\
1 + 3 mm&11.0& 8.3\\
1 + 5 mm&20.6&16.2\\
2 + 4 mm&18.9&15.8\\
1 + 2 + 3 mm&9.5&9.4\\
1.5 mm triangular&12.0&7.5\\
2 mm square& 20.0&17.1\\
2 + 3 mm square & 20.8&18.0\\
1 $\times$ 2 mm rectangular& 14.9& 12.6\\
Mixture& 11.8& 11.0
\label{table}
\end{tabular}
\end{table}

\begin{figure}
\includegraphics[width=\columnwidth]{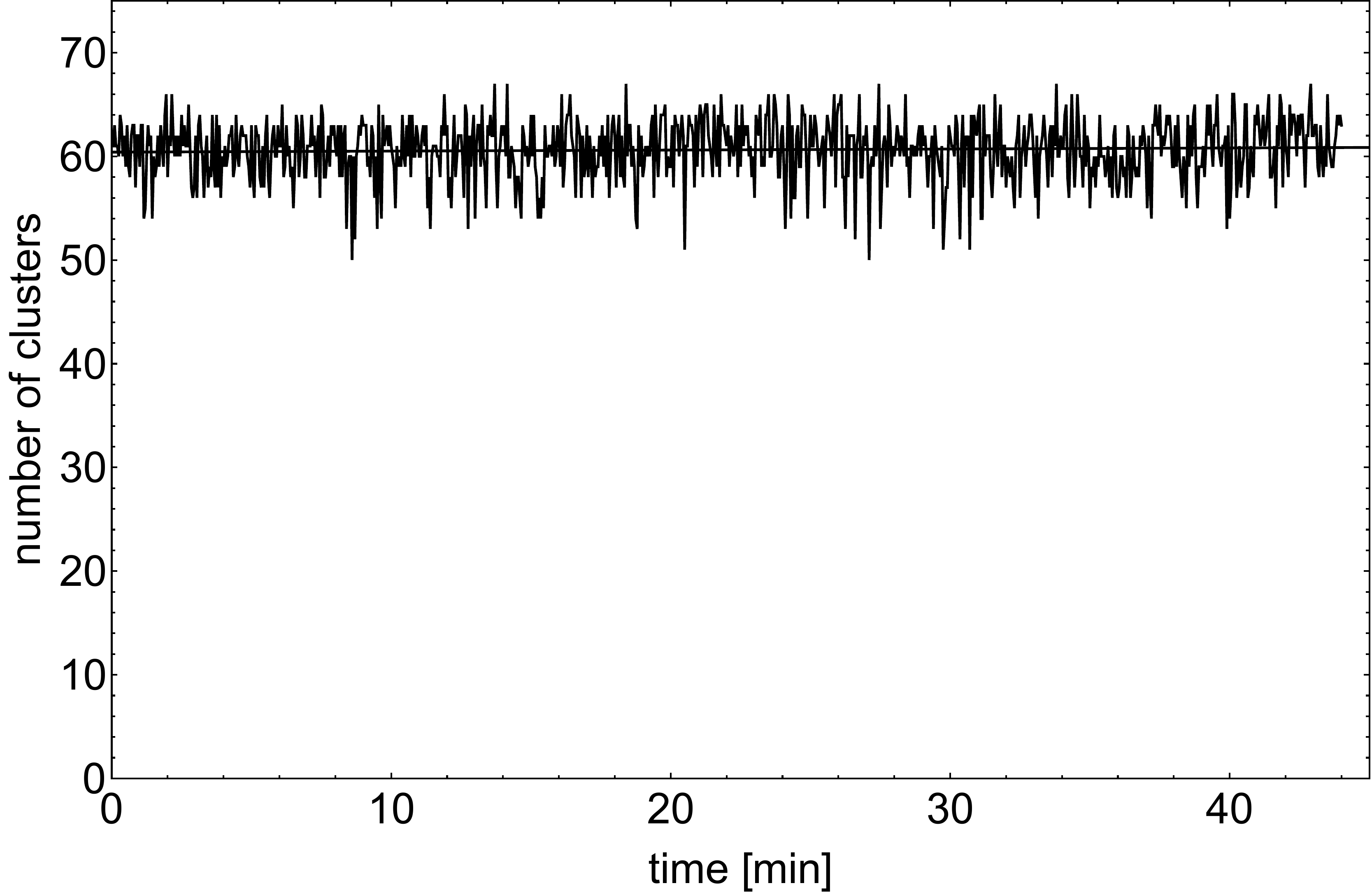}
    \caption{\label{fig.clusternumber}Number of individual aggregates over time for one example ensemble}
\end{figure}

Collision velocities in protoplanetary disks systematically vary with size, e.g., in random, turbulent motion \citep{Ormel2007}. With this in mind, at first glance, the average collision velocities in Table \ref{table} might also show a correlation with shape and size. We therefore note that there is no systematic variation. The values were just set by a random excitation. With some variation, our velocities are essentially Maxwellian distributed as is, e.g., also considered for particles in clumps formed by streaming instabilities \citep{Carrera2015}.
In contrast to earlier work by \citet{Kelling2014} we varied the shape of the aggregates and mixed aggregates of different size in an ensemble. Observations used lower spatial and temporal resolutions 
to track the ensembles. As seen above, individual collisions frequently
lead to sticking. This is also visible at lower resolution, as the clusters of aggregates from dimers to clusters of a few aggregates move together for a certain time. However, these clusters are destroyed again, eventually, as clusters interact with other clusters or aggregates in further collisions. 
In the 50 minutes observed the visual number of clusters always stayed close to the original number of individual aggregates used (Figure \ref{fig.clusternumber}). This shows that on the long-term no growth of clusters occurs.
All ensembles -- without exception -- showed this behavior.
Neither shape nor size variation promotes the stable formation of larger clusters.

\section{Conclusion}
\label{sec:conlusions}

Whatever the shape of compact dust aggregates in the studied range
of sizes and collision velocities, the result is always the same. Although individual collisions, studied in detail for cylindrical aggregates, lead to sticking between two aggregates with
a high probability, this has no effect on the long-term evolution of an aggregate ensemble. The aggregate contacts within a cluster are so weak that clusters are never stable in the given data sets (low collision velocities). In the long-term clusters are always destroyed again. This also holds for the ensembles of different shapes and sizes, although we did not quantify these in high resolution imaging.

Following \citet{Windmark2012b}, one might estimate that the probability, $p$, to obtain a large cluster consisting of $n$ aggregates would be ${p=x^n}$, where $x$ is the sticking probability. This might eventually lead to a large cluster or a lucky winner. However, this cannot be applied at the bouncing barrier as, in addition to sticking and bouncing, there is also detachment, which is a fragmentation but only undoing a previous collision. With 20\% sticking all aggregates should form dimer clusters after five collisions on average, and so on. However, even after more than 1000 collisions for each aggregate there is no stable growth. The probabilities of growing a large cluster are obviously suppressed. Currently, we cannot quantify the detachment probability from observations directly, but it prohibits the formation of larger clusters in our experiments. So at the bouncing barrier -- at least in the cases studied -- the detachment probability has to be on the same order as the sticking probability or larger.

This behavior does not rule out further growth but this requires, e.g., a seed large enough
that collision velocities go beyond meters per second where fragmentation becomes important. Still, a bouncing barrier in such scenarios among the smaller aggregates would be important or beneficial \citep{Teiser2009, Windmark2012a}.
The velocities considered for collisions by \citet{Carrera2015} for growth are in the sub-millimeter per second range. These are included in our collisions but we cannot say what the lowest limit of detaching collisions is. So we cannot rule out that growth can proceed at significantly lower velocity ranges. The aggregates used in this study are composed of silicate grains that are micrometer sized. It is well known that the grain size as well as material have major effects on sticking \citep{Okuzumi2012, Wada2013, Gundlach2015, Musiolik2016}. How variations of size and material would change the outcome of collisions is a subject for future studies. We also did not study how aggregation would approach the bouncing barrier and what the final size of aggregates would be, but concentrated on a saturated case of compact aggregates. A more self-consistent growth starting from smaller and more porous aggregates is also planned for the future.
In any case, without loss of generality, our experiments show that the bouncing barrier among small compact aggregates is very robust. 

\section*{Acknowledgements}
This work is funded by the DFG as KE 1897/1-1.

\label{lastpage}

\end{document}